\documentclass{article}

\usepackage[english]{babel}

\usepackage[letterpaper,top=2cm,bottom=2cm,left=3cm,right=3cm,marginparwidth=1.75cm]{geometry}

\usepackage{amsmath}
\usepackage{graphicx}
\usepackage[colorlinks=true, allcolors=blue]{hyperref}
\usepackage{scrextend} 

\title{\textbf{Sketch Vision: Artificial Intelligence with Sight for Imagination}

\vspace{0.25cm}

4.453 Creative Machine Learning

Demircan Tas
(tasd@mit.edu)

\date{May 4, 2023} 

}

\begin{document}
\maketitle

\section{Introduction}

\begin{figure}[h!]
  \caption{ The Origin of Painting 1786 painting by Jean-Baptiste Regnault (Museum: Museum of the History of France).}
  \includegraphics[width=0.80\textwidth]{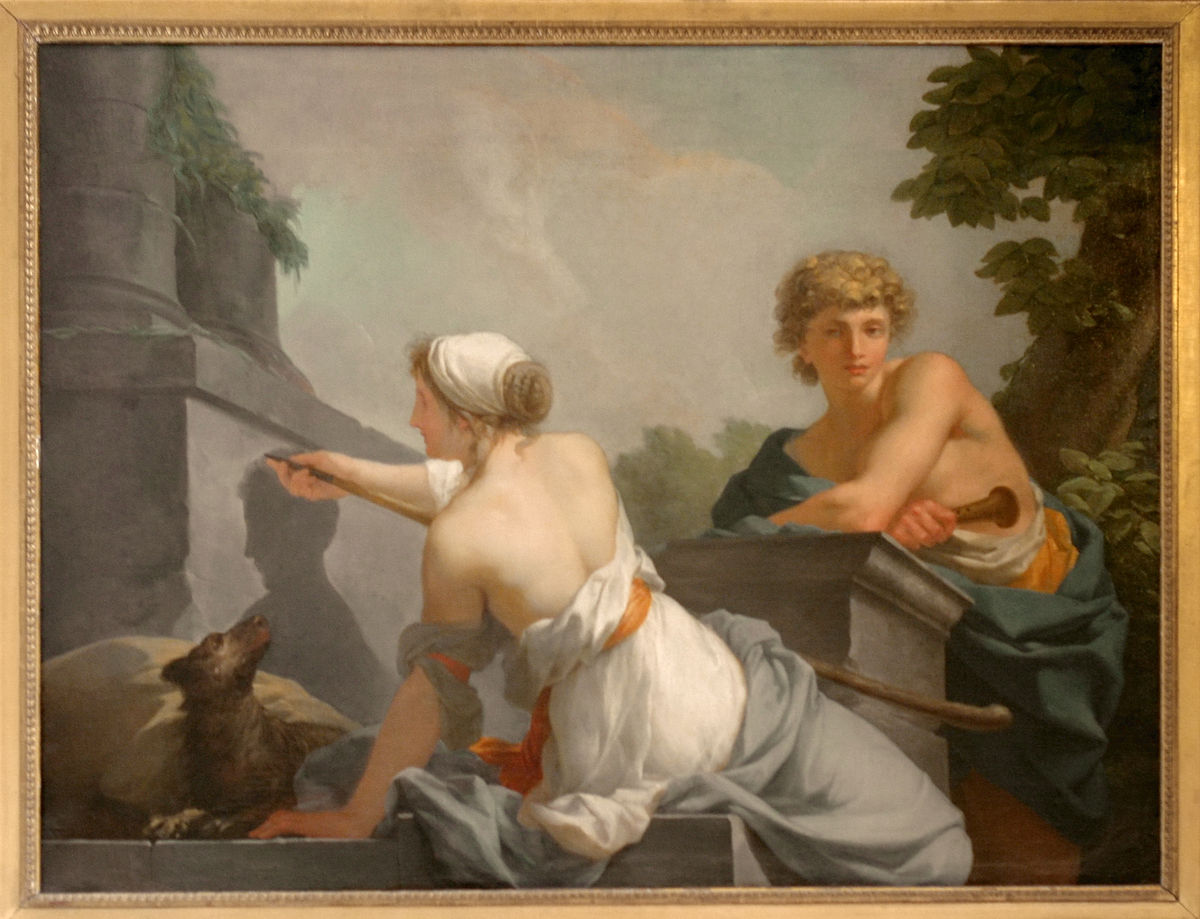} 
  \label{fig:origin}
\end{figure}

Visual design relies on seeing things in different ways, acting on them, and seeing results to act again. Parametric design tools are often not robust to design changes that result from sketching over the visualization of their output. We propose a sketch to 3d workflow as an experiment medium for evaluating neural networks and their latent spaces as a representation that is robust to overlay sketching.

Can state of the art computer vision methods assist designers in seeing? Are they limited to generative uses with predefined surfaces in the design space? Since its emergence, computation in design disciplines created an expectation of alleviating burdens of the design process, either as a smart assistant, creative partner, or a fully automated design tool \cite{aish2013first} \cite{coons1963outline}. The results while convenient in many ways, are short of the full promise \cite{steinfeld2017dreams}. Parametric design results in a paradigm of architects programming structures via code, yet programming usually follows a process that starts from the completion of traditional sketches and physical models, resulting in these new professionals being less focused on the design, and more on optimization \cite{llach2015builders}. Grammar based rules are proposed as a method of pruning the myriad of uninteresting options contained in algebras as universes of design \cite{mitchell1990logic}.

An upcoming paradigm in computation is differentiable programming, exemplified by multi-layer perceptrons, convolutional neural networks, and transformers. Machine learning has been utilized in building technology, and representation of architecture, as well as documentation. Generative AI is addressing the generation of images via prompts, and potential use for early stage design tasks.

Machine learning relies on the definition of domains similar to parametric design, followed by optimizing variables for a desired outcome based on weights acquired via training. Contemporary computer vision overcomes constraints imposed by structures by using pixel representations. Unlike an object oriented representation with defined relationships among constituents, pixels only hold values matched to coordinates. Any relationship among pixels is not defined in advance, but trained using data, in the layers of a neural network. This enables computer vision models to tackle a vast array of tasks from automated medical diagnosis, autonomous vehicles, and even generative systems of art. Computer vision can input an image, and provide a verbal definition, or do the opposite. An image can be generated from another image, and even noise, based on training data \cite{isola2017image}. The style of one image can be combined with the content of another \cite{gatys2016image} \cite{huang2017arbitrary} \cite{johnson2016perceptual}. Unsupervised computer vision is capable of detecting concepts that do not readily exist in its own structure, but inferred from data \cite{forsyth2002computer}. 

However, the output distribution in the design space of generative AI is not sufficiently stochastic 
\cite{isola2017image}. Machine learning is inherently limited to the breadth and diversity of examples included in the training set, often leading to biases, which may be perceived as a distinct style in popular examples like Dall-E and Midjourney. Moreover, generative AI tools lack the capacity for generating images with compositional qualities. These models fail when prompted with certain numbers of objects with specific spatial relationships. None of these models can generate ``two squares drawn side by side''.

\subsection{Problem Definition}

Parametric design (cite Neil Leach) creates a divide between code, and the form that is generated by it. Code, is an abstraction of the world, blind to its physical manifestation. Sensory apparatus in architecture are developed to feed pre-defined parameters of a system, enabling sensing only as part of an immutable structure.

A major shortcoming of the separation of form and the generative code appear when designers do sketch overlays. It is common in a design context to overlay tracing paper on an existing image, and sketch out design ideas. This approach is in start contrast to the parametric design paradigm, and changes made with sketches commonly lie outside of the design space defined by the parameters of the code. This fundamental difference often necessitates code revisions following design ideation.

We use the form-code duality of parametric design as a model for tackling adaptation in the context of architecture. If a three-dimensional model can be robust against sketches drawn over a rendering of it, we can extend this idea to a system adapting to unforeseen effects of nature.

Differentiable programming is a paradigm that uses neural networks as general function approximators. These functions are not explicitly defined by a programmer, but are derived from data. If the context changes, the system can be fine-tuned by additional data. This approach has been successful for a plethora of tasks in machine learning. Modern computer vision, based on this approach is a prominent domain of study with implications for design and architecture.

We propose a sketch to 3D model based on neural radiance fields (NeRF) as a function that takes two dimensional sketches, and outputs three dimensional models using NeRFs as a representation medium. Neural networks that act as the backbone of this approach are built on parameters as well. However, the parameters of a neural networks are abundant, not pre-defined, and flexible based on training data. A major property of neural networks is differentiability. Differentiable programs can be optimized to approximate any function. Differentiability, paired with a metric for evaluation, affords systems with an ability to \textit{learn} myriads of functions from nature. We aim to evaluate these qualities in an architectural design context to answer the question: Is computer vision capable of seeing like a designer? Moreover can a designer be involved in the process?

\section{Related Work}

Parametric design is based on programming each step of a design process as distinct objects, and defining chosen variables as parameters that can be manipulated after the design is completed to view alternative results based on unforeseen design constraints. This approach proves useful when the main ideas of the design are already fixed, for adapting the results to small changes in dimensions. Moreover, parameters can be optimized using evaluation metrics. Many alternative designs can be generated and evaluated against a metric, approaching design as an optimization problem in a well defined search space. Methods exist for rapid, and real-time multi-objective design optimization \cite{terzidis2006algorithmic}.

However, major design changes usually fall outside the scope of changing parameters, and require partial or a complete reconfiguration of the algorithm \cite{stiny2006shape} \cite{dossick2011messy}. Many times, the topology of a three dimensional model dictates its shape, while deformations are possible, topological change requires an application of re-topology \cite{pan2018automatic}.

\subsection{Cognition, vision, and design}

Efforts have been made to explicitly document the activities of making and design \cite{gursoy2015visualizing} \cite{knight2015making}. Design activity is carried out through the senses of a designer, creating a relationship with space where the environments acts as an extension of the person, and a memory for design computation \cite{sass2016embodied}. Among the senses of a designer, sight has the role of defining a layer of abstraction, enabling the simplification of complex states presented by the environment. Imaginary lines can use to distinguish common typologies among objects that vary in shape, but share purposes under fixed contexts \cite{minsky1988society}. Representing the vision of spaces via raster (pixel) representation is a convenient approach for enabling machine learning methods for architecture \cite{peng2017machines}.

\subsubsection{Design software disconnect}

Despite a body of research in design computation, designers and architects resort to sketches and physical processes when creative solutions are necessary. A disconnect exists between designers and computational tools. Design software depend on designers reasoning like a user, while what designers need is to see like a designer \cite{steinfeld2017dreams} \cite{sutherland1975structure}. This results in creative professionals overlaying vellum paper on computer screens or CNC milling scale models to mark desired changes with modeler's tape \cite{verlinden2005investigation}.

\section{Methodology}

\begin{figure}
    \centering
    \includegraphics[width=\textwidth]{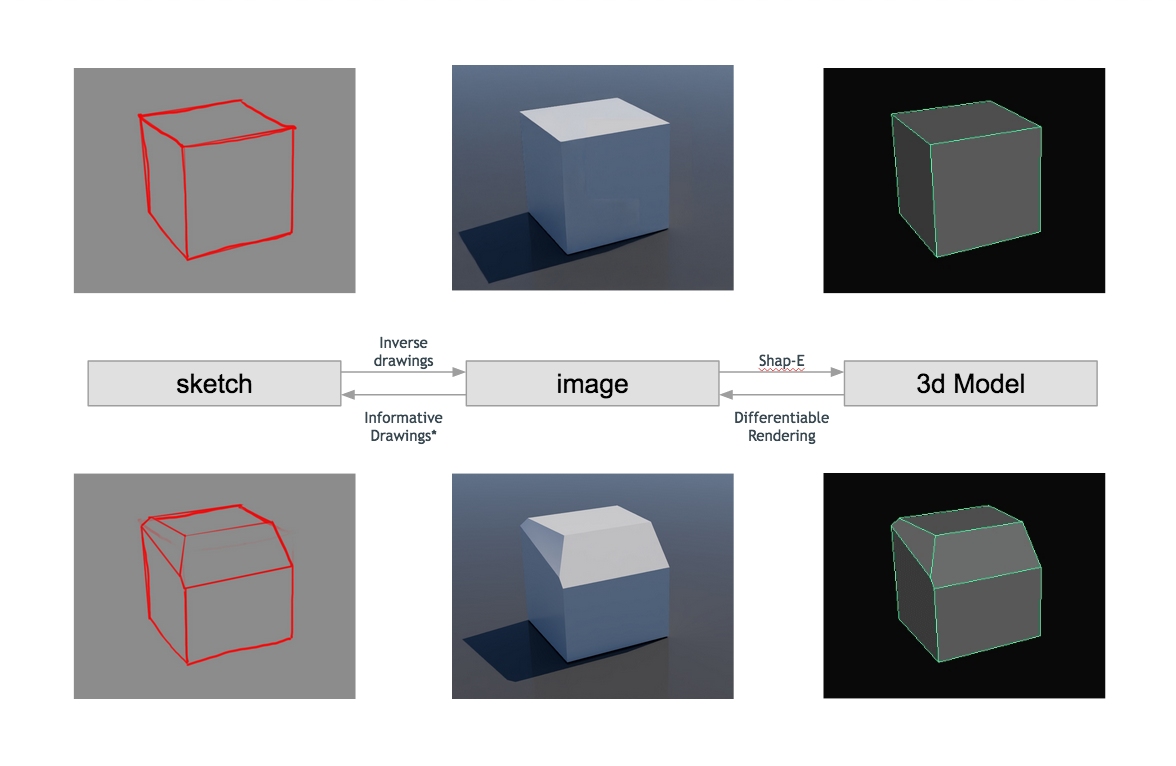}
    \caption{Diagram of Sketch Vision pipeline}
    \label{fig:pipeline}
\end{figure}

We propose a pipeline for a bidirectional, automated transition between hand drawn sketches, and three-dimensional models (see Figure \ref{fig:pipeline}). This transition is divided into two steps, using photographic images as a middle step. We train \textit{Inverse Drawings}, our own model based on \cite{isola2017image}, \cite{chan2022learning}, to go from line drawings to photographic images. Using \cite{jun2023shap}, we go from photographic images to three-dimensional models. For the reverse direction from three-dimensions to sketches, we use neural rendering to acquire photographic images from NeRFs, and \textit{Informative Drawings}\cite{chan2022learning} to go from photographic images to pseudo hand-drawn sketches. Out of the four described models for shifting modality, we use three with only minor adjustments.

For sketch to photographic image shift, we train our own model based on the reversal of the architecture proposed for \textit{Informative Drawings}. While there are existing models that go from sketch to photos using Pix2Pix\cite{isola2017image}, these models rely on image matching and adversarial loss to map sketches to images. Our proposal is based upon the idea that design sketches involve semantic and geometric qualities of three dimensional objects. To reflect this approach, we introduce additional loss metrics to the training process. By implementing \textit{CLIP}\cite{radford2021learning} into the training process, we add a semantic loss, and by including depth-maps as labels in the training, we use a pre-trained depth inferring model\cite{miangoleh2021boosting} to compute geometry loss.

\begin{figure}
    \centering
    \includegraphics[width=\textwidth]{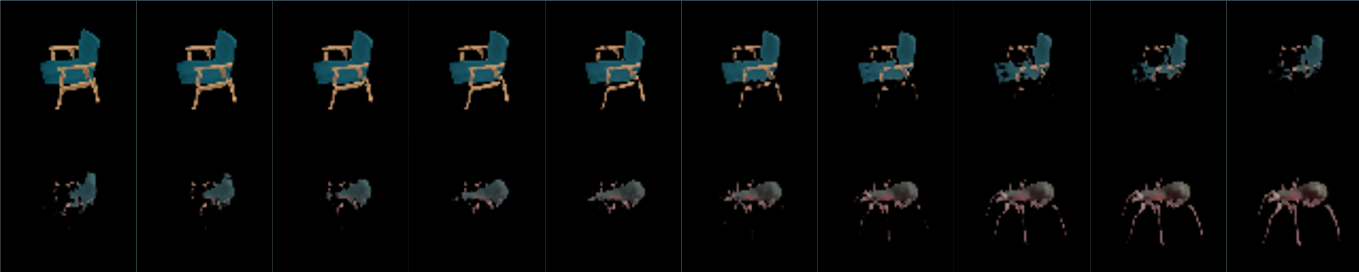}
    \caption{Linear interpolation among latent vectors for a chair and a spider}
    \label{fig:spider}
\end{figure}

\subsection{Latent Vector Interpolation}
Shap-E relies on an encoder-decoder architecture for generating three-dimensional models from images or text prompts. The encoder runs an optimization to locate a million dimensional vector for a matching image. The vector is taken into a forward pass with the decoder to acquire an implicit neural representation. By interpolating among two vectors in $n$ steps, we can pass $n$ acquired vectors into the decoder to generate intermediate objects (see Figure \ref{fig:spider}).

\subsection{Data}

\begin{figure}
    \centering
    \includegraphics[width=0.6\textwidth]{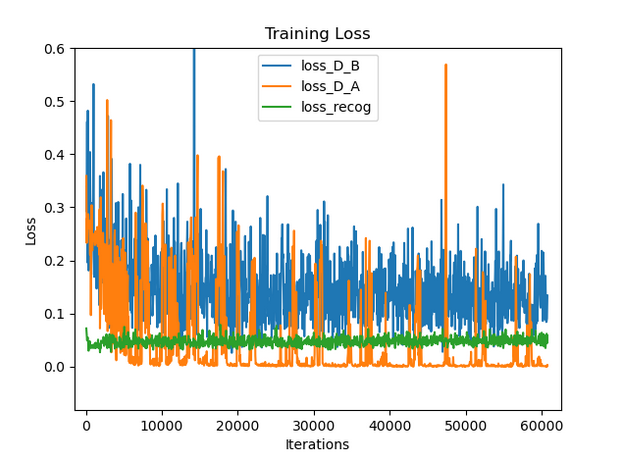}
    \caption{Plot of training losses}
    \label{fig:plot}
\end{figure}

We use 8,144 images from the Stanford Cars Dataset\cite{krause2013collecting} for training Inverse Drawings (see Figure \ref{fig:plot} for loss plot). We acquire sketches for training input, and depth maps as labels by using Informative Drawings. Moreover, we shuffle photographs from the data-set to use for computing style loss. For the second phase of our experiments, we trained the model with the same scheme, but using 8,144 images from ShapeNet Render\cite{xu2019disn} data-set. When running Informative Drawings, we used the provided pre-trained models from\cite{chan2022learning} and \cite{miangoleh2021boosting}.

\section{Results}
We present the isolated results from \textit{Inverse Drawings}, as well as the holistic results of the whole Sketch-Vision pipeline. Due to the ambiguous nature of the sketch to three-dimensions process, we provide qualitative evaluations. When trained on Stanford Cars Dataset, \textit{Inverse Drawings} exhibited strong generalization. Using design sketches of fictional wheeled vehicles as input, the output quality was similar to in domain, production cars. We observed a decrease in quality with sketches of robots and flying ships, and a greater decrease with sketches by the author. Moreover, sketches with greater accuracy tend to yield better results.

\begin{figure}
    \centering
    \includegraphics[width=\textwidth]{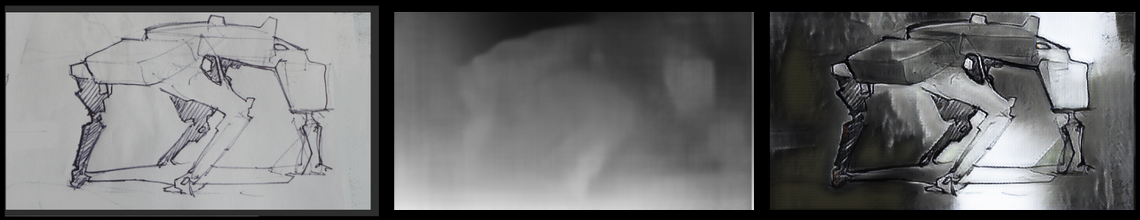}
    \caption{Results from Inverse Drawings (sketch created by author)}
    \label{fig:author}
\end{figure}

\begin{figure}
    \centering
    \includegraphics[width=\textwidth]{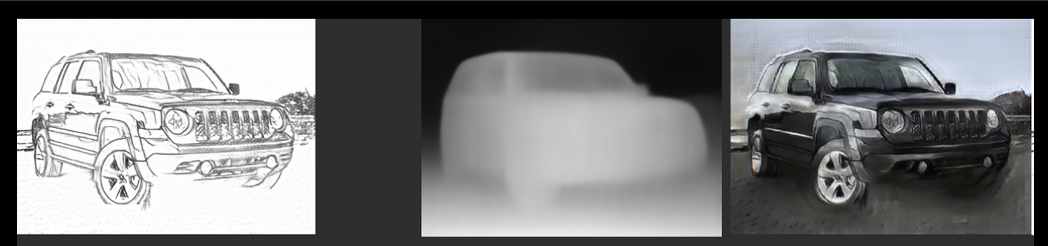}
    \caption{In-domain result from Inverse Drawings}
    \label{fig:jeep}
\end{figure}

\begin{figure}
    \centering
    \includegraphics[width=0.6\textwidth]{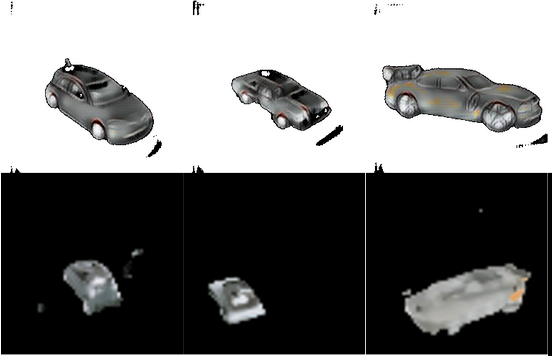}
    \caption{Shap-E with photographic input from Inverse Drawings}
    \label{fig:synth}
\end{figure}

Combining the output of \textit{Inverse Drawings} with Shap-E, we observed that the output style of Inverse Drawings was incompatible with the expected input of Shap-E. Since Shap-E was trained on synthetic data with white backgrounds, it does not generalize to input images with non-white backgrounds. By training \textit{Inverse Drawings} on ShapeNet Renders, a data-set of synthetic models rendered in Blender, we achieved a better match (Figure \ref{fig:synth}). The latter data-set lacks the diversity of the former, resulting in a lowered capacity to generalize. These constraints were not only in the domain of objects, but also in lighting conditions, camera angles, and material qualities.

\section{Conclusion}
We present \textit{Sketch-Vision}, a pipeline for automated modality shifts among hand drawn sketches, photographic images, and Neural Radiance Fields. By implementing digital images of sketches as an input, we provide a robust interaction scheme that generalizes to a breadth of use cases, from napkin sketches to digital sketch-overs. Moreover, renders or photographs from existing three-dimensional objects can be \textit{sketchified} to achieve a malleable medium for design modification.

\begin{figure}
    \centering
    \includegraphics[width=\textwidth]{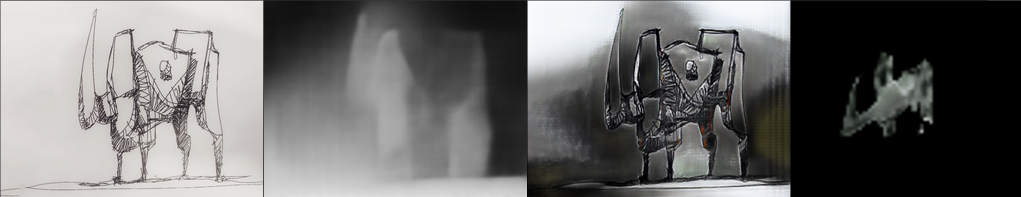}
    \caption{End-to-end example of Sketch Vision (sketch created by author)}
    \label{fig:dino-cow}
\end{figure}

\textit{Inverse Drawings} generalizes to a limited capacity when trained on rich data-sets with diverse images from the physical world, and works with other models of our pipeline to output three-dimensional representations. By combining a large model trained on a massive, yet domain-specific data-set, with our own model, we achieve an ability to steer said model with our own. As long as our training data is balanced between diversity, and the visual style of \textit{Shap-E}, we achieve representative results in three-dimensions (see Figure \ref{fig:dino-cow}).

\begin{figure}
    \centering
    \includegraphics[width=0.6\textwidth]{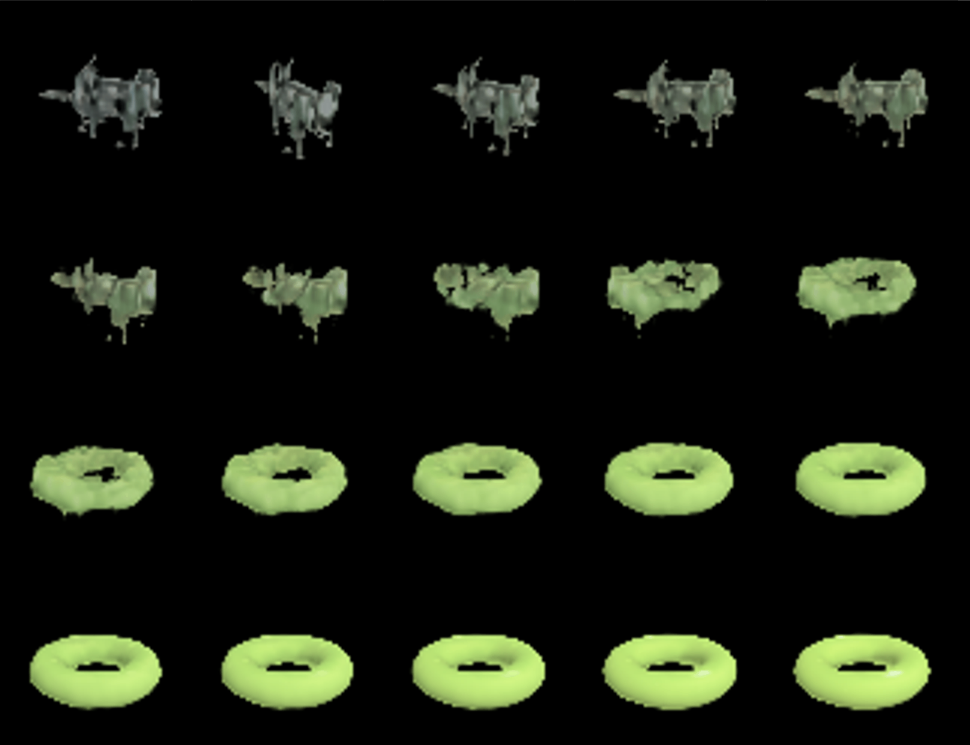}
    \caption{Linear interpolation among author's design and the image of a torus}
    \label{fig:torus}
\end{figure}

By running the pipeline with multiple sketches, or iterations of a single one, we can acquire latent vectors for each, which we can interpolate in a manner similar to parametric design (Figure \ref{fig:torus}).

\pagebreak

\bibliographystyle{acm}
\bibliography{sample}

\end{document}